\documentclass[journal=jacsat,manuscript=article]{achemso}
\usepackage[version=3]{mhchem} 
\usepackage{bm}
\usepackage[colorlinks,urlcolor=blue,anchorcolor=blue,linkcolor=blue,citecolor=blue,breaklinks=true]{hyperref}

\author{Wen-Chao Chen$^{1}$}
\author{Yuan Zhou$^{1,2,3}$}
\email{zhouyuan@nju.edu.cn}
\author{Shun-Li Yu$^{1,3}$}
\email{slyu@nju.edu.cn}
\author{Wei-Guo Yin$^{2}$}
\email{wyin@bnl.gov}
\author{Chang-De Gong$^{4,1}$}

\affiliation{$^1$National Laboratory of Solid State Microstructure, Department of
Physics, Nanjing University, Nanjing 210093, China\\
$^{2}$ Condensed Matter Physics and Materials Science Department, Brookhaven National Laboratory, Upton, New York 11973, U.S.A.\\
$^{3}$ Collaborative Innovation Center of Advanced Microstructures, Nanjing University, Nanjing 210093, China\\
$^{4}$Center for Statistical and Theoretical Condensed Matter
Physics, Zhejiang Normal University, Jinhua 321004, China}

\title{Width-tuned magnetic order oscillation on zigzag edges of honeycomb nanoribbons}

\begin{document}

\begin{abstract}
Quantum confinement and interference often generate exotic properties in nanostructures. One recent highlight is the experimental indication of a magnetic phase transition in zigzag-edged graphene nanoribbons at the critical ribbon width of about 7 nm [G. Z. Magda et al., Nature \textbf{514}, 608 (2014)]. Here we show theoretically that with further increase in the ribbon width, the magnetic correlation of the two edges can exhibit an intriguing oscillatory behavior between antiferromagnetic and ferromagnetic, driven by acquiring the positive coherence between the two edges to lower the free energy. The oscillation effect is readily tunable in applied magnetic fields. These novel properties suggest new experimental manifestation of the edge magnetic orders in graphene nanoribbons, and enhance the hopes of graphene-like spintronic nanodevices functioning at room temperature.
\end{abstract}

\textit{Introduction}---Quantum phenomena are often evident when the samples are downsized to nanometer scale due to the quantum interference effect \cite{Oka-RMP2014}. Quantum confinement realized in nanostructures thus becomes a fruitful approach to the generating and control of remarkable physical properties of matter. Among them, the possibility of finding novel magnetic properties in graphene-based nanomaterials has been of paramount interest since graphene, a single honeycomb layer of carbon atoms, was isolated from graphite and confirmed to possess extraordinary electron transport properties of massless Dirac fermions \cite{Novoselov-Nature2005,Zhang-Nature2005}.
It has been demonstrated that local magnetic moments can form on the boundary of zigzag terminated graphene nanoislands \cite{Rossier-PRL2007}, nanodisks \cite{Ezawa-PRB2007}, and nanoribbons\cite{Jiang-JCP2007,Jung-PRB2009,Hu-PRB2012,Hu-JPCC2014}.
Hence, the questions as to whether and how the boundary magnetic moments order, particularly in graphene and graphene-like ribbons such as silicene \cite{Meng-NanoLetters2013} and hafnium\cite{Li-NanoLetters2013}, have attracted much attention.

To date, the undoped parent phases of zigzag terminated graphene nanomaterials have been well studied. Density functional theory calculations predicted that the ground state of such a nanoribbon has antiferromagnetic (AF) interedge superexchange interaction \cite{Lee-PRB2005,Jiang-JCP2007}, i.e., the antiferromagnetically correlated edge (AFCE) state (total spin $S=0$). The first-principles electronic structures can be accurately reproduced in the half-filled one-orbital Hubbard model for zigzag-edged honeycomb lattices in mean-field theory \cite{Yazyev-PRL2008,Jung-PRB2009} and quantum Monte Carlo simulation \cite{Ma-PRB2016,Ma-JPCM2016}. First-principles studies of graphene triangles ($S\ne 0$) and hexagons ($S=0$) \cite{Rossier-PRL2007} further confirmed the applicability of Lieb's theorem concerning $S$ in the half-filled one-orbital Hubbard model for bipartite lattices \cite{Lieb-PRL1989}. Upon charge doping, it was found in the same model that the spin polarizations on the two ribbon edges can change from antiparallel to parallel, forming the ferromagnetic correlated edge (FMCE) state \cite{Jung-PRB2009}. The experimental indication of one AFCE-FMCE phase transition was recently reported in scanning tunnelling microscopy measurements which reveals an electronic bandgap of about 0.2$-$0.3 eV for the ribbons narrower than $7$ nm \emph{but} gapless bands for the ribbons wider than $8$ nm \cite{Magda-Nature2014}. Such a semiconducting to metallic phase transition was again accurately reproduced in the mean-field theory of the Hubbard model which found the driving force to be the AFCE-FMCE transition \cite{Magda-Nature2014}. This discovery stimulates the search for more novel effects in zigzag-edged nanoribbons and for the proper understanding of these effects.

Here, we examine the slightly doped graphene nanoribbons by studying the Hubbard model with the zigzag-edged honeycomb lattice structure. We report the finding of a robust oscillation in the edge magnetic order upon increasing the ribbon width, namely a series of alternating AFCE-FMCE and FMCE-AFCE transitions, and unveil its microscopic origin.

\begin{figure*}[tbp]
\includegraphics[width=\textwidth]{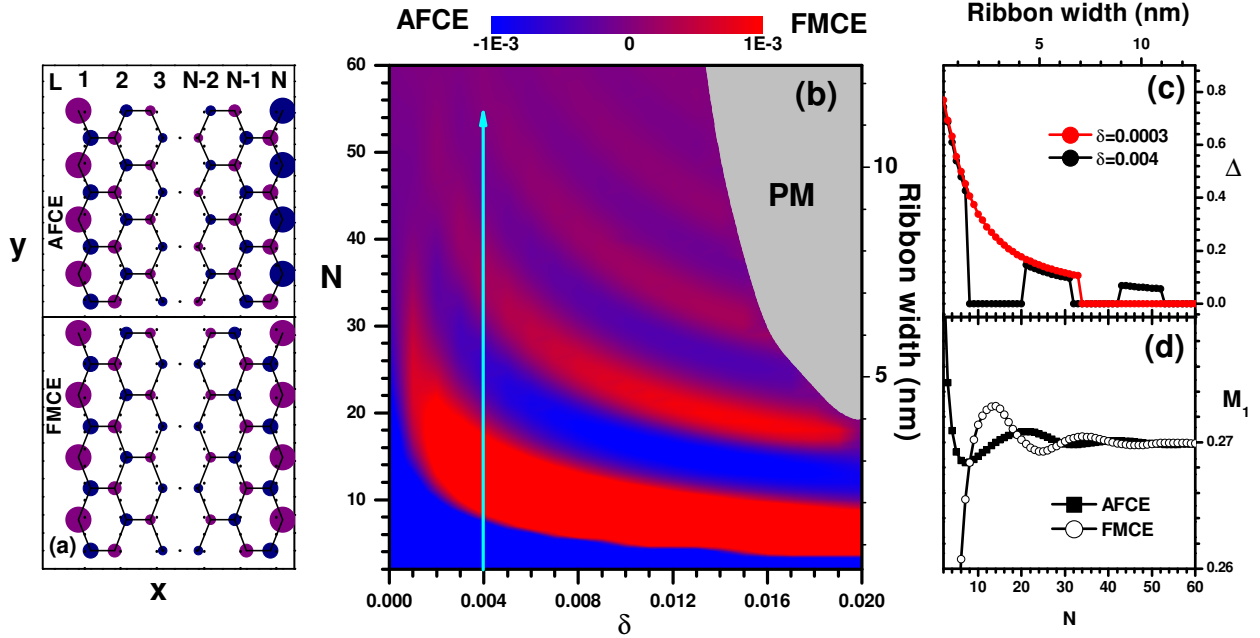} 
\caption{(a) The schematic of the AFCE and FMCE magnetic configurations in terms of the magnetization distribution in the $x$ direction (integrated along the $y$ direction. The lattice structure of zigzag-edged honeycomb nanoribbons with open boundary set in the $x$ direction (i.e., the 1st and $N$th zigzag chains are the two edges) and the periodic boundary condition set in the unconfined $y$ direction. (b) Magnetic phase diagram as a function of the ribbon width and doping level $\delta$. (c) The bandgap $\Delta$, measuring the energy difference between the lowest upper band and highest lower band, as a function of $N$ for $\delta=0.004$ and $\delta=0.0003$. (d) The strength of the edge magnetization in both the AFCE and FMCE phases for $\delta=0.004$.}
\label{F1}
\end{figure*}

\textit{Model and methods}---The lattice structure of the zigzag-edged honeycomb nanoribbons and the AFCE/FMCE states are depicted in Fig.~\ref{F1}(a). The structure is characterized by $N$ coupled zigzag chains. The ribbon width $w \approx [2+3(N/2-1)]a$ where for graphene nanoribbons $a=0.142$ nm is the carbon-carbon bond length \cite{Neto-RMP2009}. For example, $N=34$ corresponds to $w \approx7$ nm.

The single $\pi$-orbital Hubbard model, which is capable of describing the low-energy physics of graphene \cite{Magda-Nature2014,Fujita-JPSJ1996,Son-PRL2006,Pisani-PRB2007,Kobayashi-PRB2005,Rossier-PRL2007}, is described by the following Hamiltonian
\begin{equation}  \label{eq1}
H=-t\sum_{\left\langle i\text{,}j\right\rangle \sigma }c_{i\sigma }^{\dag
}c^{}_{j\sigma }+U\sum_{i}n_{i\uparrow }n_{i\downarrow}-\mu
\sum_{i}n_{i}+h_{ext}\sum_{i\sigma}\sigma n_{i\sigma},
\end{equation}
where $c_{i\sigma}$ is the electron annihilation operator with spin index $\sigma=\pm 1$ at site $i$ and $n_{i\sigma}=c_{i\sigma}^{\dagger}c^{}_{i\sigma}$ is the electron number operator. Only the nearest-neighbor hopping $t=2.6$ eV is considered in the kinetic term \cite{Jung-PRB2009}. The effective on-site Coulomb repulsion $U$ is material dependant; it is about $1.2 t$ in benzene, $2.0 t$ in silicene \cite{Schuler-PRL2013}, and $0.8-2.3 t$ in graphene \cite{Yazyev-PRL2008,Jung-PRB2009,Magda-Nature2014}. Here we consider $U=2.0t$, $1.5t$, $1.2t$ and find that our main findings remain qualitatively unchanged. $\mu$ is the chemical potential determined by the electron density $\langle n\rangle=\frac{1}{N_{T}}\sum_{i\sigma}n_{i\sigma}=1-\delta$ with $\delta < 0.02$ ($7.6\times10^{13}$ cm$^{-2}$) being the hole concentration, which can be adjusted by doping or voltage bias \cite{Geim-NatMater2007}, and $N_{T}$ the total number of lattice sites. $h_\mathrm{ext}$ denotes an external in-plane magnetic field. The temperature is fixed at $T=0.01 t$ ($\sim 300$ K). Therefore, the magnetic properties discussed here can be realized at room temperature.

The infinite system (i.e., $N\to \infty$) is metallic except for strong enough $U>U_{c}\sim 2.2t$ \cite{Fujita-JPSJ1996}, for which the ground state becomes antiferromagnetic insulating. By contrast, for zigzag-edged ribbons, spontaneous magnetization below $U_{c}$ is possible due to stronger localization on the edges \cite{Jung-PRL2009}. We compared the free energies of the paramagnetic (PM), AFCE, and FMCE states to determine the magnetic phase diagram (see Appendix I for technical details).

\textit{Results}---One of our main results, the magnetic phase diagram in terms of the ribbon width or $N$ versus the doping level $\delta < 0.02$, is shown in Fig.~\ref{F1}(b). We found a pronounced oscillation behavior of the magnetic phase in the range of $\delta=0.002-0.016$ ($0.76-6.1\times10^{13}$ cm$^{-2}$), namely the alternating stabilization of the AFCE and FMCE states upon increasing $N$, leading to the multiple first-order phase transitions. For the larger $\delta$, the magnetic state becomes more unstable and the system turns to be PM for large $N$. The period of the oscillation exhibits a strong doping dependence: it increases as $\delta$ decrease, e.g., about $2.6$ nm for $\delta=0.012$, $4.7$ nm for $\delta=0.004$, and $\infty$ at zero doping where the AFCE solution is the ground state by Lieb's theorem \cite{Lieb-PRL1989}. As shown in Figs.~\ref{S2} and~\ref{S3} in Appendix I, the predicted magnetic oscillation is robust against the on-site repulsion $U$ in the range of values widely used in literature. In particular, it remains nearly unchanged in lower doping region, although the period of oscillation extends slightly for weaker $U$.

\begin{figure}[tbp]
\vspace{-0.0in} \hspace{-0.0in} \center
\includegraphics[width=0.65\columnwidth]{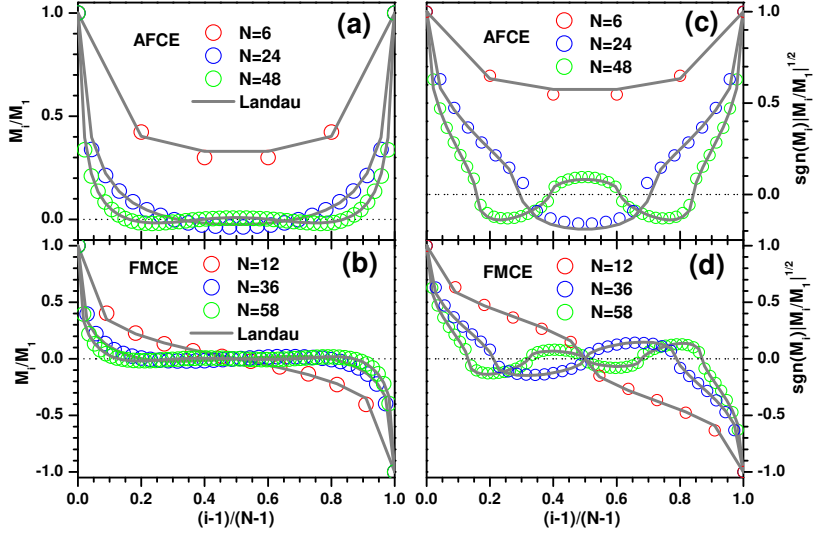} \vspace{-0.0in}
\caption{Normalized magnetization of the $i$th zigzag chain of the nanoribbon for the (a) AFCE and (b) FMCE states for $\delta=0.004$. The light gray lines result from Landau theory presented in Eq.~(\ref{E4}). To highlight the oscillations inside the ribbon, the value of $sgn(M_{i})\vert M_{i}/M_{1}\vert ^{1/2}$ is shown instead of $M_{i}/M_{1}$ in the right panels.}
\label{F2}
\end{figure}

As shown in Fig.~\ref{F1}(c), the multiple magnetic transitions manifest themselves as multiple semiconductor-metal transitions, which are characterized by the gap opening and vanishing, respectively, in tunnelling spectroscopy measurements. The experimentally observed one semiconductor-metal transition at $7$ nm effectively corresponds to our results for $\delta\approx 0.0003$.

We now look into the edge magnetic moments, since they dominate the magnetization of the system. As illustrated in Fig.~1(a), each of the $N$ zigzag chains of the ribbon contains two sublattices $A$ and $B$. Let $m_i^A$ and $m_i^B$ be the averaged magnetic moment of the $A$ and $B$ sublattices of the $i$th chain, respectively; then, $M_i=m_i^A-m_i^B$ measures the staggered magnetization of the $i$th chain. Since the outermost atoms on the left ($i=1$) and right ($i=N$) edges belong to the $A$ and $B$ sublattices, respectively, one finds from symmetry consideration that $M_1=M_N$ for the AFCE state and $M_1=-M_N$ for the FMCE state. In Fig.~\ref{F1}(d), we plot the strength of the edge magnetization, $|M_1|=|M_N|$, in both AFCE and FMCE states for $\delta=0.004$. Generally speaking, the change in the relative strength of $M_1$ of the two states coincides with the oscillation in the more stable phase [cf. black dots in Fig.~\ref{F1}(c)]. This means that the system lowers its free energy by enhancing the edge magnetization---via a reversal of edge spin correlation.

\begin{figure}[tbp]
\vspace{-0.0in} \hspace{-0.0in} \center
\includegraphics[width=0.65\columnwidth]{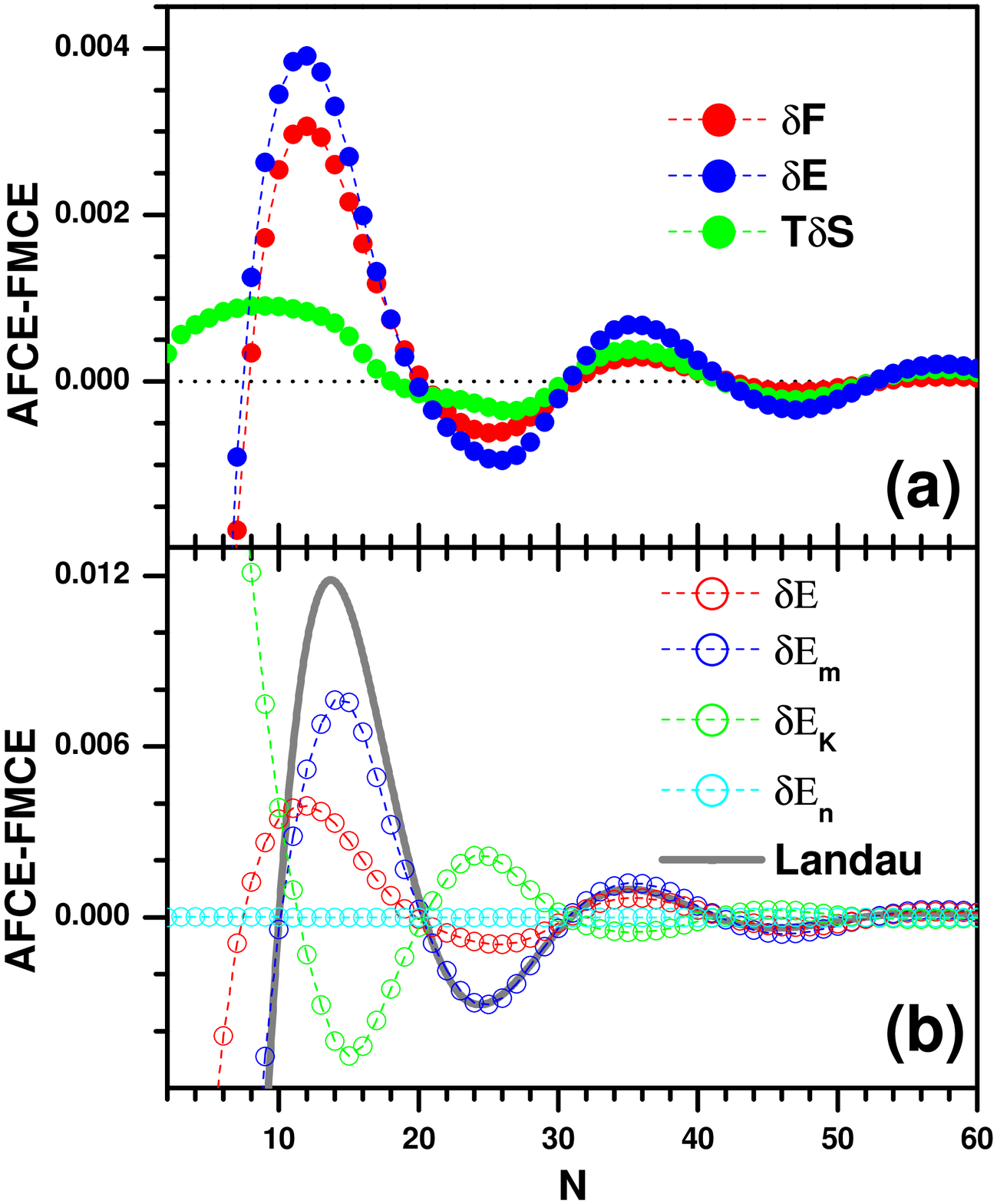} \vspace{-0.0in}
\caption{Analysis of energy difference between the AFCE and FMCE states for $\delta=0.004$. (a) Difference in the free energy $\delta F$, total energy $\delta E$, and the entropy part $T\delta S$. (b) Individual contributions from the kinetic, magnetic, and charge sectors ($\delta E_K$, $\delta E_m$, and $\delta E_n$, respectively). The light gray lines result from Landau theory presented in Eq.~(\ref{E4}).}
\label{F3}
\end{figure}

To understand how the magnetizations on the two edges correlate with each other, in Figs.~\ref{F2}(a) and \ref{F2}(b) we present the distribution of magnetic moments inside the system for the AFCE and FMCE states, respectively (see Fig.~\ref{S4} in Appendix I for more $N$-specified data). It is clear that the two edges dominate the magnetization in either case. Interestingly, the magnetization insides the ribbon is also oscillatory though weak. To clearly show this feature, we rescale the two plots in Figs.~\ref{F2}(c) and \ref{F2}(d) using the function $sgn(M_i)|M_i/M_1|^{1/2}$. From edge to edge $M_i$ changes sign $2(n-1)$ and $(2n-1)$ times respectively for the $n$th AFCE and FMCE phases, which are numbered by their appearance in Fig.~1(b) as $N$ increases for fixed $\delta$. For example, for $\delta=0.004$ and $N=48$, the ground state is the 3rd AFCE state appearing in Fig.~1(b); then, $M_i$ changes sign 4 times in Fig.~\ref{F2}(a). This behavior resembles the Friedel oscillation typically exhibited around impurities \cite{Zuo-NJP2015,Zhu-PRL2002,An-PRL2006} or standing waves in the one-dimensional quantum well, where the quantum interference plays the crucial role. The edges of the present nanoribbons can also been viewed as a kind of special impurities, yielding a similar but one-dimensional modulation along the finite-size direction. Such a modulation also resembles the oscillatory decay of the special edge states along the finite direction in the quantum spin Hall systems \cite{Lu-EPL2012}. Figs.~\ref{F2}(c) and \ref{F2}(d) manifest that the AFCE and FMCE states will emerge when the phase shift between the two edges matches $2n\pi$ and $(2n+1)\pi$, respectively, leading to the oscillatory period of about $33a$ for $\delta=0.004$, as shown in Figs.~\ref{F1} and \ref{F3}. This means that the edge magnetization is enhanced by acquiring the positive coherence with the other edge due to the quantum interference, which is realized by adjusting the edge magnetic correlations.

To gain more insights into the microscopic origin of the multiple width-dependent phase transitions, we analyzed the data for $\delta=0.004$ (the results for $0.012$ are presented in Appendix I). In Fig.~\ref{F3}(a) we show that the critical widths of the phase transitions, which are determined by the difference in free energy between the AFCE and FMCE states $\delta F=F_\mathrm{AFCE}-F_\mathrm{FMCE}$, also follow the total energy difference $\delta E=E_\mathrm{AFCE}-E_\mathrm{FMCE}$ as well as the entropy ($S$) via $T\delta S=\delta F-\delta E$ except for very narrow ribbons. To be more specific, in Fig.~\ref{F3}(b) we show the individual contributions from the kinetic, magnetic, and charge sectors ($E_K$, $E_m$, and $E_n$, respectively, defined in Appendix I). The influence of the charge imbalance ($\delta E_n$) is negligible at the low doping levels considered here. On the other hand, the kinetic energy difference $\delta E_K$ exhibits a reversal effect compared with the magnetic energy difference ($\delta E_m$), which tends to localize the electron motion. In short, the magnetic phase transitions track well the magnetic energy difference except for the very narrow ribbons. This means that we can understand the phase transitions by analyzing the physics in the magnetic sector for simplicity. Hence we present a Landau theory below.

\textit{Landau theory}---The Landau free energy that describes the spatial profile of the magnetization is given by (see Appendix II for details)
\begin{equation}
F=\int dx\left[f_{0}+\frac{1}{2}\alpha M^{2}(x)+\frac{1}{4}\beta
M^{4}(x)+\cdots\right], \label{E4}
\end{equation}%
where $\alpha<0$ and $\beta>0$ are the fitting parameters, and $M(x)$ is the staggered magnetization of the $x$th zigzag chain, which is defined for $x=i$ as $M_{i}=m_{i}^{A}-m_{i}^{B}$ \cite{Fallarino-PRB2015}. Since the magnetic momentum exhibits an oscillatory decay, it can be simulated by
\begin{eqnarray}
M^{\pm }\left( x\right) &=&M_{L}^{\pm }e^{-\gamma \sqrt{x-1}}\cos \left(
\kappa (x-1-x_{0}) \right)  \nonumber \\
&&\pm M_{R}^{\pm }e^{-\gamma \sqrt{N-x}}\cos \left(\kappa (N-x-x_{0})\right),  \label{E5}
\end{eqnarray}%
where $\gamma $ is the decay ratio and $\kappa =2\pi/\lambda $ with $\lambda $ being the oscillatory period. The superscript $+$/$-$ is for the AFCE/FMCE state. $M_{L/R}^{\pm }$ is the amplitude of edge magnetization for the left ($x=1$) and right ($x=N$) edges, which can be determined from the Euler-Lagrange equation at sufficiently large size ($N\gg 1$); In the absence of the external magnetic field, $M^{\pm}_{L}=M^{\pm}_{R}=\sqrt{\left\vert\alpha/\beta\right\vert}/\cos(\kappa x_{0})$.

As argued above, the AFCE-FMCE phase oscillations are magnetically originated, dominated by the edge magnetization due to the exponentially decay in the bulk. Therefore, a large value of $\vert M(1)\vert$ is expected to minimize the magnetic energy $E_{m}$. According to Eq.~(\ref{E5}), the edge magnetization in the AFCE state is enhanced by acquiring the positive coherence between the two edges when $n-1/4<(N-1-x_{0})/\lambda <n+1/4$ with $n$ being an integer, corresponding to the $n$\textit{th} AFCE phase. Otherwise, the edge magnetization in the FMCE state will increase. The edge magnetic correlations have to oscillate between the AFCE and FMCE states to lower the energy as the ribbon width increases. This simple phenomenological theory can accurately reproduce the above numerical results, as shown in Fig.~\ref{F2} and Fig.~\ref{F3}(b) with the parameters $x_{0}=1.8$, $\gamma=1.2$, $\lambda=22$, $\alpha=-2.3$, $\beta=31.6$ for $\delta=0.004$, except for very narrow ribbons. It is thus clear how the phase coherence due to the quantum confinement and interference between the two edges substantially control the magnetic properties of the nanoribbons.

\begin{figure}[tbp]
\vspace{-0.0in} \hspace{-0.0in} \center
\includegraphics[width=0.65\columnwidth]{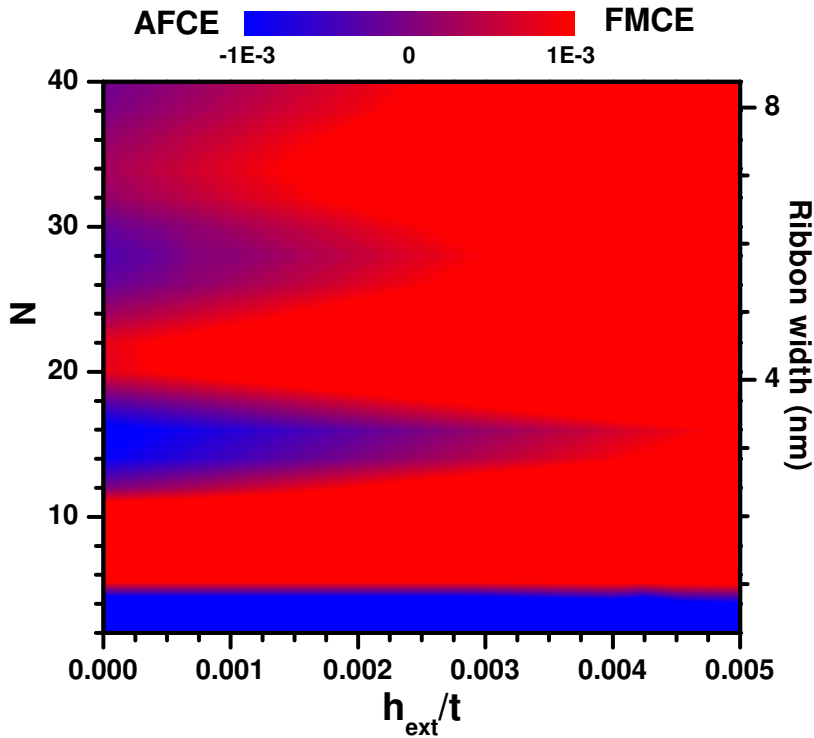} \vspace{-0.0in}
\caption{Phase diagram in an external magnetic field for $\delta=0.012$.}
\label{F4}
\end{figure}

\textit{Effects of magnetic fields}---The responses of the AFCE and FMCE states to the external magnetic field are different, since an uniform magnetic field favors the FMCE state. This provides an operational way to control the edge magnetization. The phase diagram in the presence of an in-plane external magnetic field is shown in Fig.~\ref{F4} for $\delta =0.012$. The FMCE state is strongly enhanced by the external magnetic field. It requires a significantly lower field to switch the AFCE state to the FMCE state for wider ribbons. In other words, the multiple magnetic transitions facilitate the field manipulation of the edge spin polarization. The strength of the external magnetic field can be estimated by $h_{ext}/\mu_\mathrm{B}$ with $\mu_\mathrm{B}$ being the Bohr magneton, yielding several to tens of Tesla, which is achievable experimentally.

It is noteworthy that to date it is hard to fabricate perfect zigzag edges. However, it was shown that the main effect of the edge irregularity is likely to yield a higher effective value of $U$. We also studied the magnetic phase stiffness against the Anderson-type disorders (see Appendix III). We conclude that the width-dependent oscillatory behavior of the magnetic phases are quite robust.

\textit{Conclusions}---The phenomenon of width-tuned magnetic order oscillation in the Hubbard model for zigzag-edged honeycomb nanoribbons has been unveiled. We also establish in Landau theory a simple picture of the magnetic phase oscillation, namely the positive coherence between the two edges enhances the edge spin polarizations and lower the free energy due to the quantum interference. The edge spin polarization inside the ribbon is also oscillatory, changing orientation even ($2n-2$) and odd ($2n-1$) times for the $n$th AFCE and FMCE states, respectively. We further show that the multiple magnetic transitions facilitate the field manipulation of the edge spin polarization. The oscillation effect points to new experimental manifestation of the edge magnetic orders in graphene nanoribbons. These magnetic properties are found to be quite robust against edge imperfection, operable at room temperature, and thus promising for future spintronics application.

\begin{acknowledgement}
We thank H. Q. Lin and Y. F. Wang for helpful suggestions and discussions. This work was supported by the National Nature Science Foundation of China under Contract No. 11274276 and 11674158, the Ministry of Science and Technology of China 2016YFA0300401, and the U.S. Department of Energy (DOE), Office of Basic Energy Science, under Contract No. DE-SC0012704. Y.Z. acknowledges the visiting scholarship of Brookhaven National Laboratory and the financial support of China Scholarship Council. The authors declare no competing financial interests.
\end{acknowledgement}

\section*{Appendix I. Mean-field solution of the model Hamiltonian}
Considering that the boundary condition is open in the $x$ direction and periodic in the $y$ direction, the mean-field decoupling of the Hubbard model can be transferred into the half-momentum space ($k$ for the $y$ direction) as
\begin{eqnarray}  \label{Sq2}
H &=&-t\sum_{i,k\sigma }\left( \gamma _{k}a_{i,k\sigma }^{+}b_{i,k\sigma
}+a_{i,k\sigma }^{+}b_{i-1,k\sigma }\right) +\text{h.c.}  \nonumber \\
&&+\sum_{i,k\sigma }\left(\mu _{i,\sigma}^{A}a_{i,k\sigma }^{+}a_{i,k\sigma
}+\mu _{i,\sigma}^{B}b_{i,k\sigma }^{+}b_{i,k\sigma }\right)+E_{0}\text{,}
\end{eqnarray}%
where $a$ and $b$ are the electron annihilation operators in the $A$ and $B$ sublattices, respectively. $%
\gamma _{k}=\left( 1+e^{-ik}\right) $ with the distance between the
nearest-neighbor atoms in the same sublattice being the unit. $%
E_{0}=-U\sum_{i\eta }\left( \left\langle n_{i}^{\eta }\right\rangle
^{2}/4-\left( m_{i}^{\eta }\right) ^{2}\right)$ with $\eta=A, B$. The effective chemical potential $\mu_{i,\sigma}^{\eta}=-\langle n_{i}^{\eta}\rangle/2+\sigma m_{i}^{\eta}-\mu$. The
charge density and the spin polarization at a given site are defined as $%
\left\langle n_{i}^{\eta }\right\rangle =\sum_{\sigma}\langle
n_{i\sigma}^{\eta }\rangle$ and $m_{i}^{\eta }=\sum_{\sigma}\sigma\langle
n_{i\sigma }^{\eta }\rangle/2$. The effective mean-field Hamiltonian has
been shown to yield results in good agreement with first-principles
calculations \cite{Yazyev-PRL2008,Jung-PRB2009,Rossier-PRB2008}.

\begin{figure}[tbp]
\vspace{-0.3in} \hspace{-0.0in} \center
\includegraphics[width=0.5\columnwidth]{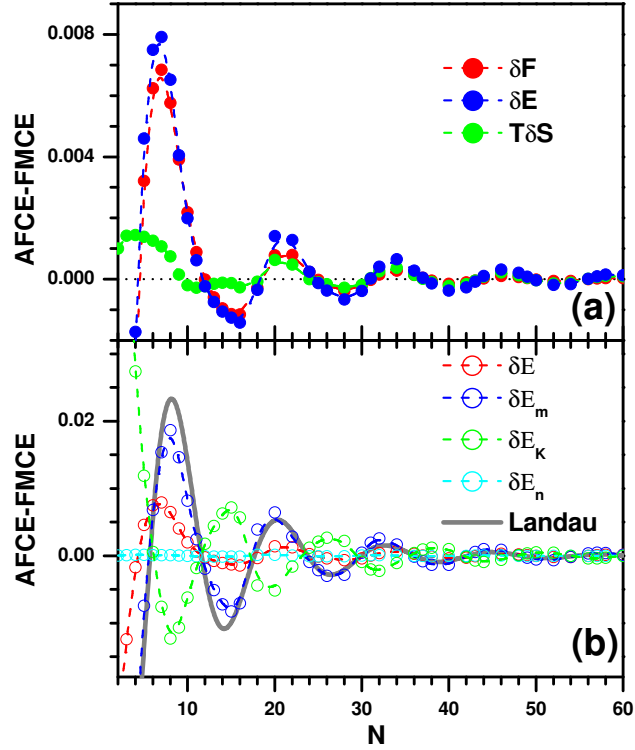} \vspace{-0.0in}
\caption{Analysis of energy difference between the AFCE and FMCE states for $\delta=0.012$ with $U=2.0t$. (a) Difference in the free energy $\delta F$, total energy $\delta E$, and the entropy part $T\delta S$. (b) Individual contributions from the kinetic, magnetic, and charge sectors ($\delta E_K$, $\delta E_m$, and $\delta E_n$, respectively). The light gray lines result from Landau theory presented below with fitting parameter $x_{0}=0.5$, $\gamma=1.1$, $\lambda=12.3$, $\alpha=-2.8$, and $\beta=49.1$.}
\label{S1}
\end{figure}

The free energy of the system is given by
\begin{equation}\label{Sq3}
F=-\frac{k_{B}T}{N_{y}}\sum_{k\sigma }\sum_{\nu =1}^{2N}\ln \left(
1+e^{-E_{k\sigma }^{\nu}/k_{B}T}\right)+\mu\sum_{\eta }\sum_{i =1}^{N}\langle n_{i }^{\eta }\rangle-E_{0},
\end{equation}
where $E^{\nu}_{k\sigma}$ is the eigenvalue of Eq.~(\ref{Sq2}) and $N_{y}$ is the number of $k$ points. The temperature $T$ is fixed at $0.01t$. Therefore, the magnetic properties discussed in present paper can be applied at the room temperature. The total energy $E$ is
\begin{equation}\label{Sq4}
E=\langle H \rangle=-\frac{\partial}{\partial \beta}\ln Z=\frac{1}{N_{y}}\sum_{k\sigma}\sum_{\nu=1}^{2N}E_{k\sigma}^{\nu}f_{k\sigma}^{\nu},
\end{equation}
where $Z$ is the partition function, $\beta=1/k_{B}T$, and $f_{k\sigma}^{\nu}=1/(1+e^{\beta E_{k\sigma}^{\nu}})$ is the Fermi distribution.
The total energy can be further divided into three parts, i.e. $E=E_{K}+E_{m}+E_{n}$ with $E_{K}=-t\langle \sum_{\left\langle i\text{,}j\right\rangle \sigma }c_{i\sigma }^{\dag
}c^{}_{j\sigma } \rangle$, $E_{m}=-U\sum_{\eta}\sum_{i=1}^{N} (m_{i}^{\eta})^{2}$, and $E_{n}=\frac{U}{4}\sum_{\eta }\sum_{i =1}^{N}\left\langle n_{i}^{\eta }\right\rangle
^{2}$ being the energy from kinetic energy, magnetization, and charge density, respectively. In Fig.~\ref{S1} we show the relative energy for $\delta=0.012$, whose oscillatory features look qualitatively similar to those of Fig.~3 in the main text for $\delta=0.004$. The AFCE-FMCE phase transition determined by the total free energy tracks well with the energy from magnetization.

\begin{figure}[tbp]
\vspace{-0.3in} \hspace{-0.0in} \center
\includegraphics[width=0.8\columnwidth]{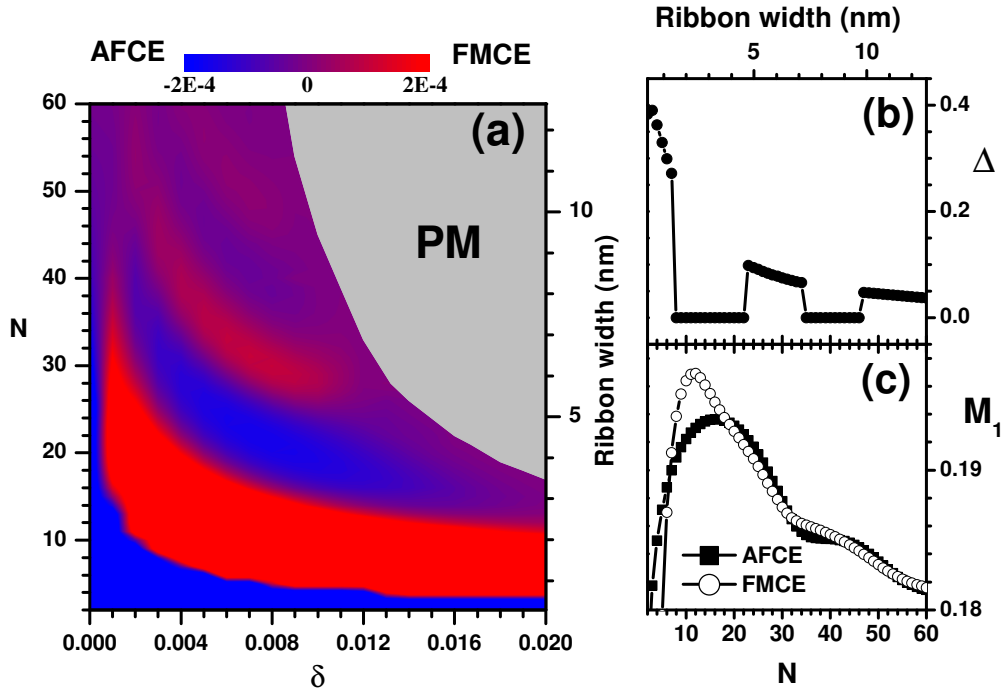} \vspace{-0.0in}
\caption{(a) Magnetic phase diagram as a function of the ribbon width and doping level $\delta$ with $U=1.5t$. (b) Bandgap $\Delta$, measuring the energy difference between the lowest upper band and highest lower band, as a function of $N$ for $\delta=0.004$. (c) Strength of the edge magnetization in both the AFCE and FMCE phases for $\delta=0.004$.}
\label{S2}
\end{figure}

\begin{figure}[!bp]
\vspace{-0.2in} \hspace{-0.0in} \center
\includegraphics[width=0.6\columnwidth]{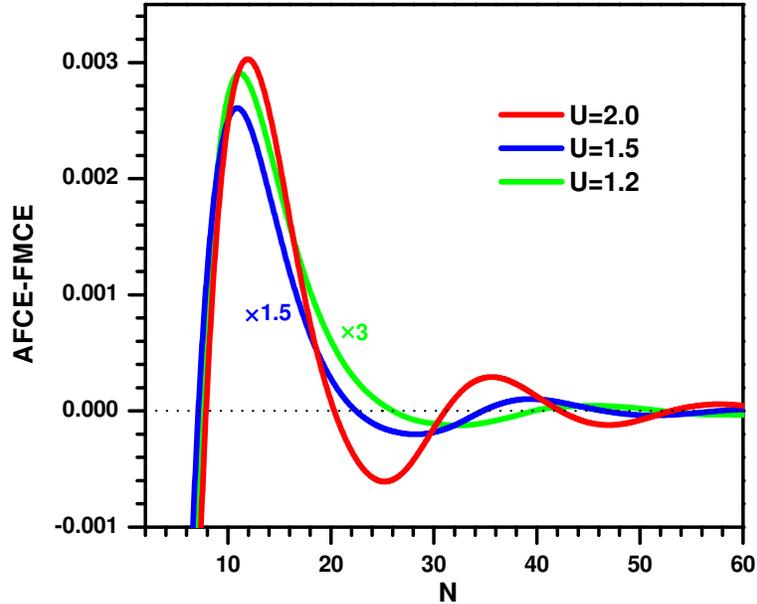} \vspace{-0.0in}
\caption{Difference in the free energy between the AFCE and FMCE states for $\delta=0.004$ with various on-site repulsion $U$. The data for $U=1.5$ and $U=1.2$ have been enlarged by a factor for comparison.}
\label{S3}
\end{figure}

The magnetic phase oscillations are robust against the on-site Coulomb repulsion $U$ adopted in various literatures. We plot the phase diagram with a weaker on-site repulsion $U=1.5t$ in Fig.~\ref{S2}. Compared with the stronger repulsion with $U=2.0t$ presented in the main text, the magnetic phase oscillations remain but with the expanded paramagnetic phase. Especially, the resultant phase transitions change little for lower dopings below $\delta=0.008$ though the period of oscillation increases slightly. In Fig.~\ref{S3} we further show the difference in the free energy between the AFCE and FMCE states for different on-site repulsion $U$ at the fixed doping $\delta=0.004$. The phase oscillation remains robust even with weak enough $U=1.2t$.

\begin{figure}[tbp]
\vspace{-0.3in} \hspace{-0.0in} \center
\includegraphics[width=0.8\columnwidth]{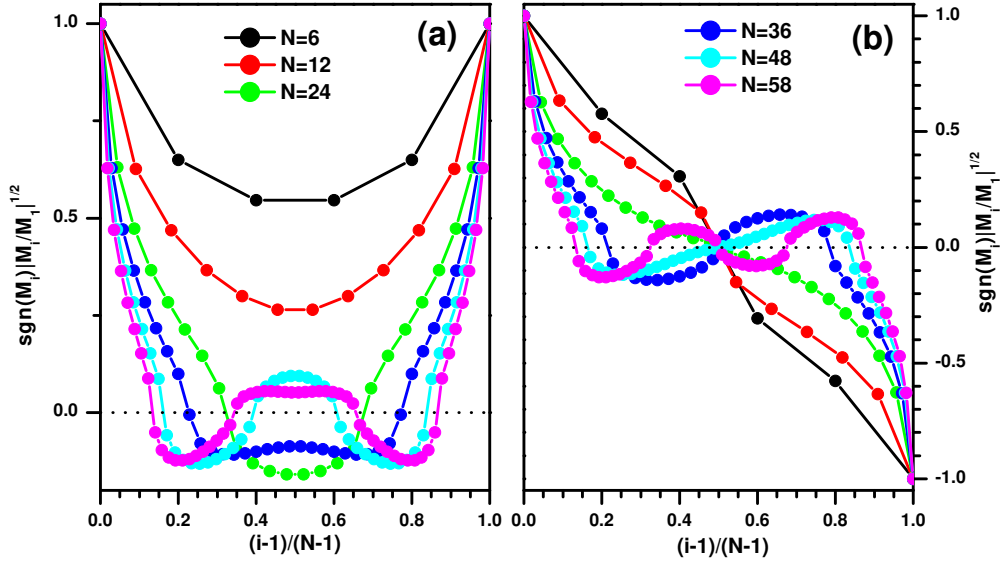} \vspace{-0.0in}
\caption{Distributions of magnetization of the \emph{i}th zigzag chain of nanoribbon for (a) AFCE and (b) FMCE states at fixed $\delta=0.004$ and $U=2.0t$. The data has been renormalized by the magnitude along the edge to highlight the oscillations inside the bulk. The magnetization of \emph{i}th zigzag chain is defined in the main text.}
\label{S4}
\end{figure}

In Fig.~\ref{S4}, we show the distributions of magnetization inside the nanoribbon for both AFCE and FMCE states for more values of $N$ than in Fig.~2 of the main text. Evident oscillation of magnetization can be found in the bulk. The oscillatory period is about $22$ ($33$a with a the carbon-carbon bond length) for $\delta=0.004$ and $U=2.0t$ as shown in the main text. To gain positive coherence from the opposite edge, the magnetic correlation between the two opposite edges has to alternate between AFCE and FMCE when the ribbon width changes. This is particularly evident for large widths. For example $N=36$, the edge magnetization acquires the positive coherence from another edge in FMCE state while negative coherence from another edge in AFCE state.

\section*{Appendix II. Landau theory description for AFCE-FMCE phase transition}
As mentioned in the main text, the stability of the magnetic phase is dominated by the distribution magnetization, our starting point is the
Landau theory with the single order parameter $M(x)$ as
\begin{equation}
F=\int dx\left[ f_{0}+\frac{1}{2}\alpha M^{2}\left( x\right) +\frac{1}{4}%
\beta M^{4}\left( x\right) -h M\left( x\right) +\cdots \right] \text{,}
\end{equation}
where $\alpha <0$, $\beta>0$ are fitting parameters, $h$ is the effective external magnetic field. $M(x)$ is the magnetization of the \emph{x}th zigzag chain, which is defined as $M_{i}=m_{i}^{A}-m_{i}^{B}$ for $x=i$ ($i=1$, $2$, $\cdots$, $N$) to account for the antiferromagnetic correlations between the nearest neighbor sublattice $A$ and $B$\cite{Fallarino-PRB2015}. The magnetic momentum follows an oscillatory decay from the boundaries and can be well simulated by%
\begin{eqnarray}
M^{\pm }\left( x\right) &=&M_{L}^{\pm }e^{-\gamma \sqrt{x-1}}\cos \left( \kappa
(x-1-x_{0})\right) \nonumber \\
&&\pm M_{R}^{\pm }e^{-\gamma \sqrt{N-x}}\cos \left(
\kappa \left(N-x-x_{0}\right)\right) \text{,}
\end{eqnarray}%
where $M^{\pm}_{L}\cos(\kappa x_{0})$ and $M^{\pm}_{R}\cos(\kappa x_{0})$  is the
order parameter at the left ($x=1$) and right edge ($x=N$) at sufficiently large size ($N\gg 1$), respectively. $\kappa x_{0}$ is
an initial phase introduced to well simulate our numerical data. $\kappa
=2\pi /\lambda $ with $\lambda $ the oscillatory period. The superscript $%
+/-$ corresponds to the AFCE/FMCE state.

The Euler-Lagrange equation at the two edges ($x=1$ and $x=N$) satisfies
\begin{equation}
\alpha M(x)+\beta M^{3}(x)-h=0\text{,}
\end{equation}
where the superscript $\pm $ has been neglected.
\bigskip

\textit{In the absence of the external magnetic field}---The Euler-Lagrange
equation at the two edges reduces to $M^{\pm }\left( 1\right) =\sqrt{%
\left\vert \alpha /\beta \right\vert }$ and $M^{\pm }\left( N\right) =\pm
\sqrt{\left\vert \alpha /\beta \right\vert }$, which requires $M_{L}^{\pm
}=M_{R}^{\pm }=M_{1}^{\pm }$. The value of $M_{1}^{\pm }$ can be evaluated at the limit of $N\rightarrow
\infty $, yielding
\begin{equation}
M_{1}^{+}=M_{1}^{-}=M_{1}=\sqrt{\left\vert \alpha /\beta \right\vert }/\cos
(\kappa x_{0})\text{,}
\end{equation}
in which the two edges are no longer correlated with each other and
therefore the difference between AFCE and FMCE disappears. In the present
case, since $M^{2}\left( x\right) \ll 1$, the dominating contribution comes
from the first term in the Landau free energy, especially
the magnetization near the edge due to exponentially decay departing from the edges.
To minimizing the total free energy, strong edge spin polarization $m(1)$ and $m(N)$ are expected.
At the edge ($x=1$)
\begin{equation}
M^{\pm}(1)=\sqrt{\left\vert\frac{\alpha}{\beta}\right\vert}\left[1\pm e^{-\gamma \sqrt{N-1}}\cos \left(
\kappa (N-1-x_{0})\right)\right].
\end{equation}
When $2n\pi-\pi/2<\kappa (N-1-x_{0})<2n\pi+\pi/2$, $M^{+}(1)>M^{-}(1)$, otherwise, $M^{+}(1)<M^{-}(1)$. The amplitude of the edge magnetization $M^{+}(1)$ and $M^{-}(1)$ is alternatively dominant upon the numbers of the zigzag chains $N$. The edge magnetization acquires the positive coherence by adjusting the edge magnetic correlations, which naturally generates the AFCE-FMCE phase oscillations. The simple Landau theory description is in good agreement with the numerical calculations, manifesting its validity. Therefore, the physics behind the phase oscillations is the quantum confinement (finite size $N$) and quantum interference (positive coherence).
\bigskip

\textit{In the presence of the external magnetic field}---The inversion antisymmetry in AFCE state is broken under the external magnetic field, leading to $M_{L}\neq M_{R}$. We consider the infinite case ($N\rightarrow
\infty $) to show this discrepancy. The induced Euler-Lagrange equation in
this situation at the two edges is (the superscript $+$ has been neglected
for simplicity) $\alpha M_{L}\cos(\kappa x_{0})+\beta \left( M_{L}\cos
(\kappa x_{0})\right)^{3}-h=0$ at left edge ($x=1$), and $\alpha \left(
M_{R}\cos (\kappa x_{0})\right) +\beta \left( M_{R}\cos(\kappa x_{0})\right)
^{3}+h=0$ at right edge ($x=N$). Using the Cardano formula for one variable
cubic equation, we have%
\begin{equation}
x_{2}=\omega \left( -q+\left( q^{2}+p^{3}\right) ^{\frac{1}{2}}\right) ^{\frac{1}{3}}+\omega
^{\ast }\left( -q-\left( q^{2}+p^{3}\right) ^{\frac{1}{2}}\right) ^{\frac{1}{3}}\text{,}
\end{equation}
where $p=\beta /2\alpha $, $q=\pm h/3\alpha $ ($+$ for the right, and $-$ for
the left edge), and $\omega =(-1+\sqrt{3}i)/2$. Usually, the external magnetic field
is $3\sim4$ orders weaker than the spontaneous magnetization, we expand the above
equation to the first order of the external magnetic field $h$, yielding
\begin{equation}
M_{L/R}\cos (\kappa x_{0})=\sqrt{\left\vert \frac{\beta }{\alpha }\right\vert } \pm%
\frac{h}{2\beta }\text{.}
\end{equation}
Therefore, the edge magnetization is enhanced at one edge and weakened at another edge due to the breaking of the inversion antisymmetry in the AFCE states under the external magnetic field.

In comparison, the inversion symmetry is preserved in the FMCE states under the external magnetic field, which means $M_{L}=M_{R}=M$.  Similar
relation $M\cos(\kappa x_{0})=\sqrt{\left\vert \frac{\beta }{\alpha }%
\right\vert }+\frac{h}{2\beta }$ is subsequently obtained according to the Euler-Lagrange equation. The edge magnetization is enhanced at both edges.
Due to the different response of the edge magnetization in the AFCE and FMCE state, the FMCE is more energetically favorable under the external
magnetic field.

\section*{Appendix III. Disorder effects}
\begin{figure}[hbp]
\vspace{-0.0in} \hspace{-0.0in} \center
\includegraphics[width=0.5\columnwidth]{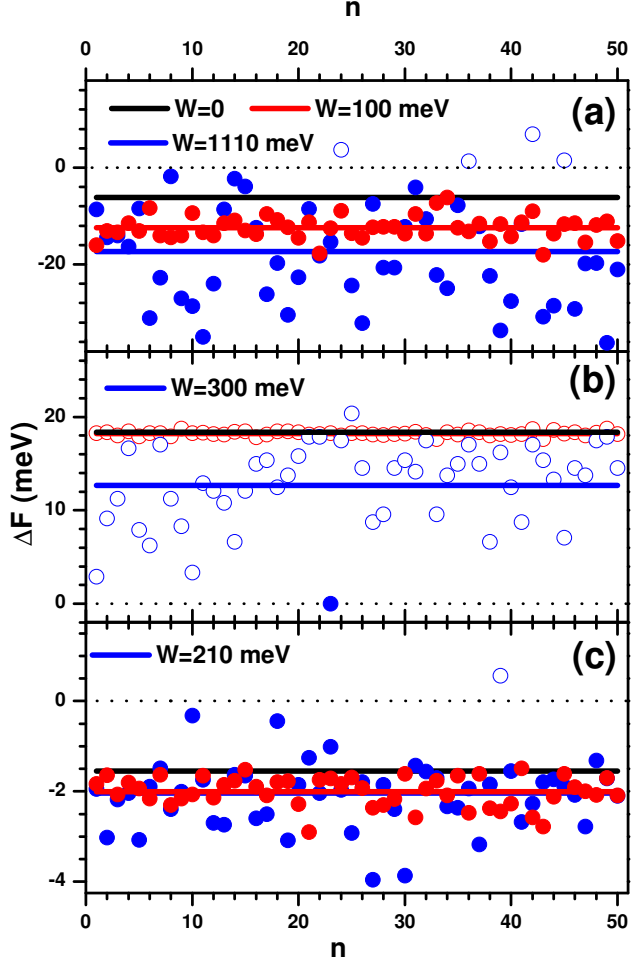} \vspace{-0.0in}
\caption{Free energy difference between the AFCE and FMCE state $\Delta F=F_{AFCE}-F_{FMCE}$ with disorders. The black dotted lines is the phase boundary of the AFCE and FMCE states ($\Delta F=0$). Red circles are for the moderate strength of disorders with $W=100 meV$. The blue circles are for the critical strength of the disorders, where the occasional AFCE-FMCE instability occurs. The solid, and hollow circles energetically favor the AFCE, and FMCE state, respectively. The solid lines (red/blue) are the averages for the respective disorders. The solid black lines are for $W=0$. $\delta=0.012$ and $U=2.0t$ and $t=2.6$ eV is adoped. (a) The $1$\emph{st} AFCE state with $N=4$; (b) The $1$\emph{st} FMCE state with $N=8$; (c) The $2$\emph{nd} AFCE state with $N=16$.}
\label{S5}
\end{figure}

We consider the non-magnetic disorders along the edges $H_{W}=\sum_{\sigma i\in edges}W_{i}n_{i\sigma}$, where $W_{i}$ is the strength of the disorder at the given site randomly distributed in the interval $[-W/2,W/2]$ with $W$ the strength of the disorders. We solve the Hamiltonian in the real-space with the periodic boundary condition $N_{y}=24$ along the infinite $y$-direction due to the broken translation symmetry along $y$-direction. In fact, the results presented here are insensitive to the large enough $N_{y}$. For simplicity, only the disorders at the edges with inversion symmetry are considered.

The influence of the edge disorders on the antiferromagnetic correlated edges (AFCE) and ferromagnetic correlated edges (FMCE) states is displayed in Fig.~\ref{S5} with $50$ times random disorders. At the doping $\delta=0.012$, the red circles with moderate strength of the edge disorders $W=100$ meV ($t=2.6$ eV \cite{Jung-PRB2009}) well locate in the respective phase for the $1$-\emph{st} AFCE ($N=4$), $1$-\emph{st} FMCE ($N=8$), and $2$-\emph{nd} AFCE state ($N=16$). Therefore, the magnetic phases are quite robust at least for the low-\emph{th} magnetic states. When the strength of the disorders is enhanced to a critical value, the magnetic phase may be unstable. For the $1$-\emph{st} AFCE state, the FMCE occasionally has lower energy for strong enough disorder $W_{c}=1100$ meV (blue hollow circles in Fig.~\ref{S5}(a)). This critical strength depends on the number of the zigzag chains $N$, it is about $300$ meV, $210$ meV, for the $1$-\emph{st} FMCE ($N=8$), and $2$-\emph{nd} AFCE ($N=16$) state, respectively. Interestingly, the average effect of the disorders seems to stabilize the AFCE states while weaken the FMCE states. This is probably due to that the scattering of the edge disorders will open a gap in the metallic FMCE state.

\bibliography{ref}

\end{document}